\documentclass[12pt,centertags,showpacs,preprintnumbers,amsmath,amssymb]{revtex4}
\usepackage[cp1251]{inputenc}
\usepackage[english]{babel}

\usepackage{graphicx}
\usepackage{epsfig}
\usepackage{dcolumn}
\usepackage{bm}
\begin{document}
\newcommand{\cn}{\mathop{\rm cn}\nolimits}
\newcommand{\tr}{\mathop{\rm tr}\nolimits}
\newcommand{\sn}{\mathop{\rm sn}\nolimits}
\newcommand{\Tr}{\mathop{\rm Tr}\nolimits}
\newcommand{\diag}{\mathop{\rm diag}\nolimits}
\newcommand{\e}{\mathop{\rm e}\nolimits}
\newcommand{\ts}{\textstyle}
\newcommand{\ds}{\displaystyle}
\newcommand{\ints}{\int\limits}
\newcommand{\nn}{\nonumber\\}
\newcommand{\eps}{\varepsilon}
\arraycolsep=1.8pt
\newcommand{\Det}{\mathop{\rm Det}\nolimits}
\font\eightrm=cmr8
\arraycolsep1.5pt

\title{ \bf Competition of color ferromagnetic and superconductive
  states in a quark-gluon system } 
\author{D.~Ebert}
\affiliation{Institut f\"ur Physik,
Humboldt-Universit\"at zu Berlin, D-12489 Berlin, Germany}

\author{V.~Ch.~Zhukovsky}

\author{O.\,V.\,Tarasov}

\affiliation{ Physical Department, Moscow State University,  119992,
Moscow, Russia}

\date{\today}

\begin{abstract}
{The possibility of color ferromagnetism in an $SU(2)$ gauge field model is
investigated. The conditions allowing a stable color ferromagnetic state of
the quark system in the chromomagnetic field occupying small domains
are considered. 
A phase transition between this 
state and the color 
superconducting states is considered. The effect of finite temperature
is analyzed.}
\end{abstract}

\maketitle
\renewcommand{\thefootnote}{\arabic{footnote}}
\setcounter{footnote}{0}
\setcounter{page}{1}
\section*{ Introduction}
Nonperturbative effects of non-abelian gauge theories take place in
the infrared region.  Among such nonperturbative effects are the existence
of the {\it QCD} vacuum with gluon and quark condensates, chiral
symmetry breaking, confinement 
and the hadronization process. They can only be studied by approximate
methods and in the framework of various effective models.
For instance, one of the possibilities to approximately describe the gluon
condensate is to introduce background color fields of certain
configurations (see, e.g., \cite{5,6}). 

Another example of nonperturbative problems is the physics of light
mesons that can be described by effective
four-fermion models such
as the Nambu--Jona--Lasinio (NJL) quark model, which was successfully
used to
implement the ideas of dynamical chiral
symmetry breaking ($D\chi SB$) and bosonization (see
e.g. \cite{EbPerv},\cite {7} 
and references therein; for a review of (2+1)-dimensional four-quark
effective models see \cite {echaya}).

In the framework of four--fermion models, it was shown that a
constant magnetic field  \cite {10} induces the dynamical chiral symmetry
breaking ($D\chi SB$),
as well as the fermion mass generation, even under conditions when
the interaction between fermions is weak.
This phenomenon of magnetic catalysis  was explained basing upon
the idea of effective reduction of space dimensionality in the
presence of a strong external magnetic field \cite {11} (see also
paper \cite {12} and references therein). It was also demonstrated
that a strong chromomagnetic field catalyzes
$D\chi SB$ \cite {13}. As was shown in \cite {zheb}, this effect
can be understood in the framework of the dimensional reduction
mechanism as well, and it does not depend on the particular form
of the constant chromomagnetic field configuration.

One of the solutions of the Yang-Mills equations that can serve 
as a model for the gluon condensate is a constant
chromomagnetic field. Its role was demonstrated in 
\cite{savv,Matinyan}.  In these papers, the authors calculated the
one-loop effective potential for a constant chromomagnetic field  $B$
and they demonstrated 
that it reaches its minimum at a nonvanishing
value of  $B$.

However  this simple analytical 
model of the gluon condensate with a uniform chromomagnetic field $B=%
\mbox{const}$ (the so-called 
``colour ferromagnetic state'') \cite{savv,Matinyan} 
suffered from an instability \cite{nielsen}.  This problem later
has been studied in a number of papers. In particular, various methods
were proposed to stabilize the solution by introducing 
a nonzero charged component of the gauge field.  
Various attempts have also been made to improve this  
by assuming
a certain domain-like \cite{HBN} or non-abelian
structures \cite{AC} of  the condensate  field. Possibilities of
thermalizing the system 
\cite{skalozubbordag,skalozub} and introducing a condensate of the
time component of the gauge field $A_0 \neq 0$ were also considered
\cite{starinets,5}.  

The principal difficulty in finding a stable color ferrromagnetic
state is that a local minimum of the action
can not be obtained, since the corresponding field configuration proved
to be spatially inhomogeneous. 
In order to circumvent this difficulty,  the method 
earlier applied in analyzing the quantum Hall effect
\cite{Hall} was used in \cite{iwaz1,iwaz2}.
It was demonstrated that a spatially
homogeneous state of the gluon field can be obtained by
effectively reducing the dimensions of the problem from  $D=(3+1)$ to
$D=(2+1)$ and employing for gluons the technique used in
\cite{Hall,Semenoff} for the description
of the quantum Hall effect in   
a $(2+1)$-dimensional Fermi system. The
resulting gauge configuration breaks the color symmetry and the
arising color charge can be compensated by 
fermions (quarks) in the system interacting with the color field. It should be
mentioned that in this case the quark density must be higher than in
the hadronic phase of matter. 

At the same time, in the system with high fermion density, another 
nonperturbative phenomenon 
can take place, i.e., the color 
superconductivity 
(CSC) (see, e.g.\cite {alf} -- \cite{Shovkovy}).
Briefly, the mechanism of CSC can be explained as follows. Quarks in 
a dense medium become asymptotically free, their interaction is rather
weak and attraction may arise between them. Components with longer 
interaction range are screened. As a result, quarks of different colors
and flavors with opposite momenta may compose Cooper pairs, as
electrons do in the superconducting metal, thus
leading to the energy gain.  One may expect that,
similar to the case of the quark condensate, the process  of
diquark condensation can be catalyzed by intensive external
(vacuum) gauge fields. For a (2+1)- dimensional model, this was
recently discussed in \cite {ebklim}.
It is clear
that  a medium with the density $\sim 
10^{14}g/cm^3$ needed
for color superconductivity to arise 
cannot be produced in
the laboratory. This dense media may be found, e.g., in compact
stars \cite{Astr1,Astr2}. 

In the present publication, we investigate color ferromagnetism (CFM) and color
superconductivity as two mutually excluding possible phases in an 
$SU(2)$ gauge field model. Our aim is to study in more detail the effective
dimensional reduction in the gluon sector leading to stabilization of
the ferromagnetic state, and also to consider for the role of
fermions in stabilizing the system and their contributions to
thermodynamical quantities in the color ferromagnetic phase. We
analyse the ferromagnetic phase and compare the  energy gains due
to either of possible effects: creation of color ferromagnetism and
color superconductivity. 

The paper is organized as follows. Section~1 contains the 
investigation of the gluon sector. Section~2 contains the
discussion of the quark sector. In Sections~3 and~4, we consider the
phase transitions at zero and finite temperatures respectively,   
and Section~5 contains conclusions. 

\section{The effective dimensional reduction in the gluon sector}

Let us consider the gluon Lagrangian in the Yang-Mills theory with the
$SU(2)_c$ color group for the field $A^a_{\mu}\,\, (a=1,2,3)$ in the 
(3+1)-dimensional space-time 
\begin{equation} 
{\cal L}_g=-\frac 14(F_{\mu \nu }^a)^2-\frac 1{2\xi }(\overline{D}^{ab}_\mu 
Q^b_\mu )^2+\overline{\chi }_a(\overline{D}^2)_{ab}\chi _b.
\end{equation}
Here $A_\mu ^a=\overline{A}_\mu ^a +Q_\mu ^a$, $\overline{A}_\mu ^a$ is the 
background field, $Q_\mu ^a$ are quantum fluctuations of the gluon 
field, $\overline{D}_\mu ^{ab}=\delta ^{ab}\partial _\mu-g\epsilon^{abc}\overline{A}_\mu ^c$ is 
the covariant derivative in the background field, 
$\chi, \overline{\chi}$ are ghost fields and $(\overline{D}%
^2)_{ab}=\overline{D}_\mu ^{ac}\overline{D}_\mu ^{cb}$, $g$ denotes
the gluon coupling constant (the Feynman  
gauge $\xi =1$ will be adopted in what follows). Moreover, summation
over repeated color indices $a =1,2,3$; and Lorentz indices
$\mu = 0,1,2,3$ is implied. The above Lagrangian can be rewritten in the form demonstrating
interaction of the colored (``charged'') vector field with $a=1,2$ and
a ``neutral'', $a=3$, i.e., ``electromagnetic'' field:
\begin{equation}
{\cal L}_g=-\frac{1}{4}f_{\mu\nu}^2-\frac{1}{2}|(D_{\mu}W_{\nu}-D_{\nu}W_{\mu})|^2-igf_{\mu\nu}W^+_{\mu}W_{\nu}
+\frac{g^2}{4}(W^+_\mu W_\nu-W^+_\nu W_\mu)^2,
\end{equation}
with the following notations adopted:
$A_{\mu}= A^3_{\mu}$ is the ``neutral electromagnetic'' field,
$f_{\mu\nu}=\partial_{\mu}A_{\nu}-\partial_{\nu}A_{\mu}$ is the
corresponding field tensor;  
$W_{\mu}=\frac{1}{\sqrt{2}}(A^1_{\mu}+iA^2_{\mu})$ is the
``charged'' field, $D_{\mu}=\partial_{\mu}+igA_{\mu}$. We omitted the
ghost contribution in the above lagrangian, having in mind that it
will cancel the unphysical degrees of freedom in subsequent
calculations. 

Let us assume that the background field is a constant abelian
chromomagnetic field directed along the $x_3$ axis, playing the role of
the ``electromagnetic'' field in the above equation. Choosing an
asymmetric gauge for this field, we can 
write 
\begin{equation}
\begin{array}{l}
{\overline A}_\mu ^a\,=\,\delta^a_3 {\overline A}_{\mu}, \\
\label{13}
{\overline A}_{\mu}\,=\,\delta_{\mu2}x_1B, \,\, \overline{f}_{12}=-\overline{f}_{21}=B.
\end{array}
\end{equation}
In the linear approximation,  the field equation for quantum
fluctuations of the
color field $W_{\mu}$ in the background  chromomagnetic
field $\overline A_{\mu}$ has the form:
\begin{equation}
\overline{D}_{\nu}^2 W^{\mu}\,-ig{\overline f}^{\mu\nu}W_\nu\,=\,0.
\end{equation}
We may write the following stationary solutions for two physical
degrees of freedom of the fluctuation field $W_{1,2}$ interacting with
the chromomagnetic field $B$:
$$
W_{1,2}=\mbox{e}^{-i \varepsilon x_0+ik_3x_3+i k_2 x_2}(u_{n})_{1,2}(x_1-k_2/g
B),
$$ 
where the functions $(u_{n})_{1,2}(x_1-k_2/gB)$ are
expressed in terms of the known wave functions of the harmonic oscillator.
The energy spectrum has the form:
\begin{equation}
\varepsilon^2=k_3^2+2gB(n+\frac{1}{2})\,+\,2\sigma gB, \label{tah}
\end{equation}
where $n=0,1,2,...$ is the Landau quantum number. It is clear that the
energy might become imaginary, when the spin projection is $\sigma=-1$, 
the Landau quantum number is $n=0$ and the longitudinal
momentum projection is restricted by  $k_3^2<gB$ (the so called tachyonic mode). 
This
behavior of the solution demonstrates that the linear approximation for
solving the equation for this particular mode is not valid
\cite{nielsen}.
Therefore,  for this unstable tachyonic mode,
which is denoted by the field $\varphi$, 
\begin{equation}
W^1=\frac {1+i}{\sqrt 2}\varphi,\,\,\,W^2=\frac {1-i}{\sqrt 2}\varphi,
\end{equation}
the exact nonlinear Lagrangian
\begin{equation}
L_{\rm{ tach}}=|D_{\mu}\varphi|^2+2g
B|\varphi|^2-\frac{g^2}{2}|\varphi|^4, \label{L}
\end{equation}
should be considered.
Unfortunately, the procedure of
finding a nontrivial  uniform vacuum state $\varphi \ne 0$ independent of
$x$, as it was done in the case of  
the Higgs model, is not applicable  here due to the presence of
$A_{\mu}(x)$ in the covariant 
derivatives $D_{\mu}=\partial_{\mu}+igA_{\mu}(x)$. In
\cite{iwaz1,iwaz2}, a new 
approach was proposed to achieve the stable vacuum state. 
There, the solution
with $k_3=0$, 
which is  uniform in the $x_3$ direction,
was considered.
This choice leads to stabilization of the
solution, if the effective dimensional reduction $D=(3+1) \to D=(2+1)$
is implemented.  

It should be emphasized that the  solution found in
\cite{iwaz1} is stable only with respect to (2+1)-dimensional perturbations. 
However, the true vacuum state should  be stable with respect to
(3+1)-dimensional perturbations as well, i.e., against 
non-uniform perturbations along the $x_3$ axis parallel to the
chromomagnetic field direction. In fact, this is the condition that there should be
no tachyonic modes with nonzero momenta along this axis,  $k_3\ne 0$. 

In order to find this condition, let us consider the unstable modes in
more detail. It follows from  (\ref{tah}) that unstable modes appear
when $n=0,$ and the sign of the gluon spin projection is negative, while the
longitudinal momentum is comparatively small, i.e., 
$k_3^2< g B$. In order to exclude tachyonic modes with  $k_3\ne 0$,
the extention of the chromomagnetic field along 
the $x_3$ axis  $L_3$ should be finite. In this case, inplying a periodicity condition
for the ``charged'' field along this axis
$W^{1,2}(x_3=0)=W^{1,2}(x_3=L_3)$, the momentum becomes discrete:
$k_3=2\pi n_3/L_3$, where $n_3=0, \pm 1,\pm 2,\dots$. Thus, in order to
exclude tachyonic modes with $|k_3|>0$, one should demand that the
lowest non-uniform modes, with $n_3=\pm 1$, should already have real
energy,  i.e., that they were not tachyonic, and this 
supplies a restriction on the maximum 
extension $L_3$ of the chromomagnetic
field $B$ along the $x_3$ axis, 
\begin{equation}
L_3 \le L_3^{max}=\frac {2\pi}{\sqrt {g B}} .
\label{maxfield}
\end{equation}
This limitation provides a physical meaning for the choice of the
solution uniform along the $x_3$ axis, which implies a 
physically reasonable restriction, i.e., the 
field should exist only inside certain domains with finite dimensions.
In other words, when the length $L_3$ (the $x_3$ dimension of the
domain) is fixed, the inequality (\ref{maxfield}) may be considered as
a limitation for the 
maximum value of the stable chromomagnetic field inside the domain
\begin{equation}
g B\le g B_0=(\frac{2\pi}{L_3^{max}})^2.
\label{Blength}
\end{equation}
 Thus, we should not consider 
a solution, which is uniform in the whole space,
but rather a single domain, 
inside which the field is constant in magnitude and in direction. 
This limitation for its maximum 
extension and the boundary conditions
imply that the domain is surrounded by other domains with fields
inside having different
strength and orientation in configuration and group spaces, restoring
the symmetries of the whole system. In fact, the above limitation for
the $L_3$ extension of the chromomagnetic field domain implies effective dimensional
reduction for the unstable mode that results in its stabilization. As
it was mentioned above, we restrict ourself by the simplest assumption of 
periodicity of
the solutions along the $x_3$ axis. The
detailed investigation of the 
structure of domains and boundary effects that inevitably occure, when
domains are put together, is out of the scope of the present article.

\section{Quark sector}
The gauge configuration
that consists of a 
constant chromomagnetic field $B$ and a uniform  gluon condensate
$W^{1,2} \ne 0$ is
color charged.  Following
\cite{iwaz1}, one has to
introduce fermions with finite density into the system that will
interact with the gauge 
field configuration and preserve its
color neutrality.  
In the case of 
larger quark matter density, 
an effective model
describing the
interaction of quarks should be employed. Let us first give
some  arguments for  the choice 
of the QCD-motivated extended Nambu-Jona-Lasinio effective model of
quarks introduced below. 
To this end, consider two-flavor QCD with nonzero chemical
potential $\mu$ and 
color group 
$SU_c(2)$. In the previous section, we decomposed the gluon field $A_\nu^a(x)$
into a condensate background (``external'') field $\overline{A}_\nu^a(x)$ and
the 
quantum fluctuations $Q_\nu^a(x)$ around it, i.e. 
$ A_\nu^a(x)=\overline{ A}_\nu^a(x)+Q_\nu^a(x).$
By integrating in the generating functional of QCD over
quantum fluctuations $Q_\nu^a(x)$ and further ``approximating'' the 
nonperturbative gluon propagator by a $\delta-$function,
one arrives at an effective local chiral four-quark interaction of
the
NJL type describing low energy hadron physics 
in the presence of a gluon condensate. Finally, by performing a
Fierz transformation of the interaction term, one obtains 
a four-fermionic model with $(\bar q q)$--and $(q q)$--interactions
and an external condensate field ${\overline A}_\mu^a(x)$
of the color group $SU_c(2)$  given by the following
Lagrangian
\begin{eqnarray}
 {\cal L}_q&=&\bar q[\gamma^\nu(i\partial_\nu+g \overline{A}_\nu^a(x)\frac{\sigma^a}2)
+  \mu\gamma^0]q+\frac{G_1}{2N_c}[(\bar qq)^2+(\bar qi\gamma^5\vec
  \tau q)^2]+\nonumber\\
  &+&\frac{G_2}{N_c}[i\bar q_c\varepsilon
\epsilon       \gamma^5 q]
  [i\bar q\varepsilon
\epsilon\gamma^5 q_c].
  \label{x1}
\end{eqnarray}

In (\ref{x1}), $q_c=C\bar
q^t$, $\bar q_c=q^t C$ are charge-conjugated spinors,
and $C=i\gamma^2\gamma^0$ 
is the charge conjugation matrix ($t$ denotes
the transposition operation),
$\vec \tau\equiv (\tau^{1},
\tau^{2},\tau^3)$ are Pauli
matrices in the flavor space; 
$(\varepsilon)^{ik}\equiv\varepsilon^{ik}$,
$(\epsilon)^{\alpha\beta}\equiv\epsilon^{\alpha\beta}$
are totally antisymmetric tensors in the flavor and color spaces,
respectively. 
Clearly, the Lagrangian (\ref{x1}) is invariant under
the chiral $SU(2)_L\times SU(2)_R$ and color $SU_c(2)$ groups.

As it was mentioned above, the  physical vacuum of $QCD$ may be
interpreted as a region splitted into an
infinite number of domains with macroscopic extension.
Inside each such domain, there can be excited a homogeneous background 
chromomagnetic field  \cite{nielsen}, which generates a nonzero gluon condensate
$\langle FF\rangle\ne0$. Averaging over all domains results in
restoration of 
the color as well as Lorentz
symmetries
\footnote{Strictly speaking, our
following
calculations refer to some given macroscopic domain.  The obtained
results turn out to depend on color and rotational (Lorentz)
invariant quantities
only, and are independent of the concrete domain.}.

 Upon performing the usual
bosonization procedure
\cite{EbPerv},\cite {7}
and introducing meson and diquark fields $\sigma,\,\pi$
and $\delta,\,\delta^{*}$, the four-quark terms are replaced by
Yukawa interactions of quarks with these fields, and the Lagrangian
takes the following form
(our notations refer to four--dimensional Euclidean space with
$it=x_4$)
\footnote{
We consider $\gamma-$matrices in the $4-$dimensional Euclidean space
with the metric tensor $g_{\mu\nu}=\diag (-1, -1, -1, -1)$, and the
relation between the Euclidean and Minkowski time
$x_{(E)}^0=ix_{(M)}^0
$: $ \gamma_{(E)}^0=i\gamma_{(M)}^0, \,
\gamma_{(E)}^k=\gamma_{(M)}^k.$ In what follows we denote the
Euclidean Dirac matrices as $\gamma_{\mu}$, suppressing the subscript
$(E).$ They have the following basic properties $\gamma_{\mu}^+= -
\gamma_{\mu},\,\{ \gamma_{\mu},\gamma_{\nu}\}=- 2\delta_{\mu\nu}.$
The charge conjugation operation for Dirac spinors is defined as $
\psi_c(x)\quad =\quad C\left(\psi(x)^+\right)^t $ with $
C\gamma_{\mu}^tC^{-1}=-\gamma_{\mu}.$ We choose the standard
representation for the Dirac matrices (see \cite{Rho}). The
$\gamma_5$
has the following properties:
$\{\gamma^{\mu},\gamma_5\}=0,\quad\gamma_5^+=\gamma_5^t=\gamma_5.$
Hence, one finds for the charge-conjugation matrix:
$C=\gamma^0\gamma^2,\quad C^+=C^{-1}=C^t=-C.$}:
\begin{eqnarray}
{\cal L}_q &=&-\bar q(i\gamma_\nu\nabla_\nu
+i\mu\gamma_0+\sigma+i\gamma^5\vec
\tau\vec\pi)q-\frac{1}{4G}(\sigma^2+\vec \pi^2)-
\frac{1}{4G_1}\delta^{*}\delta-\nn
&-&\delta^{*}[iq^tC\varepsilon\epsilon\gamma^5 q]
-\delta[i\bar q \varepsilon\epsilon\gamma^5 C\bar q^t],
\label{1}
\end{eqnarray}
where $\nabla _\mu
=\partial_{\mu}-ig\overline A_{\mu}^a\sigma_a/2$ is the covariant
derivative of quark fields in the background field.

In order to investigate
possible phase transitions in the quark matter in the framework of the
initial model (\ref{1}), we evaluate the
path integral over meson and diquark fields 
by using the saddle point approximation,
neglecting field fluctuations around the
mean-field (classical) values $<\sigma>=\sigma_{0}=0$,
$<\pi>=\pi_{0}=0$
\footnote{The vanishing of the pion mean-field $<\pi>=\pi_{0}=0$ is here related to
the assumed parity conservation of the ground state. In what follows
we consider only 
the diquark condensate, neglecting 
the possible existence of a 
quark condensate $<\sigma>=\sigma_{0}=0$.}
and $<\delta>=\Delta_{0}$, $<\delta^{*}>=\Delta^{*}_{0}$. 

Within this approximation, we obtain the quark contribution to
the partition function
\begin{eqnarray}
Z_q=\exp W_{E}
=\int dqd\bar{q} \exp \left[\int d^4x {\cal L}_q\right],
\label{1k}
\end{eqnarray}
where
\begin{eqnarray}
{\cal L}_q =-\bar q(i\gamma_\nu\nabla_\nu)q
-\Delta^{*}_0[iq^tC\varepsilon\epsilon^3\gamma^5 q]
-\Delta_0[i\bar q \varepsilon\epsilon^3\gamma^5 C\bar q^t],
\end{eqnarray}
with $W_{E}$ being the Euclidean effective action, 
and ${\cal L}_q$ the quark Lagrangian. 

The partition function (\ref{1k}) is calculated in the standard way
(for
details see, \cite{ebklim}).
In principle, the gap $\Delta_0$ is complex. However,
the partition function
is real and depends only on the module squared of the gap. Its phase
characterizes just the degeneracy of the vacuum and may be
set here equal to zero. In this sense, it is understood that
the following equations are expressed directly in terms of the
module $|\Delta_{0}|$, i.e.
\begin{equation}
2\Delta_0 \rightarrow |2\Delta_0|\,=\,\Delta.
\label{module}
\end{equation}
In the present study, the background field is assumed constant and homogeneous,
${\overline F}_{\mu\nu}^a~=~const$. Then the Dirac equation 
\begin{equation} 
\left(\gamma_{\mu} \nabla_{\mu}\right)\psi=0 
\label{dir}
\end{equation}
for a quark with flavor $i$ has stationary
solutions $\psi_{k,i}$ with the energy spectrum $\varepsilon_{k,i},$
where $k$ stands for the quantum numbers of the quark in the
background
field. In this case we arrive at the following Euclidean effective
action:  \begin{equation}
W_E=\frac 12\int \frac{dp_4}{2\pi}
\sum_{k,i,\kappa}\log
\left(p_4^2+\Delta^2+(\varepsilon_{k,i}-\kappa\mu)^2\right).
\label{3}
\end{equation}
Here, $\kappa=\pm1$ corresponds to charge conjugate contributions of
quarks
with color indices
$\alpha~=~1,2$ (included in the quantum number $k$) and the spectrum
$\varepsilon_{k,i},$ moving in the background color field
${\overline F}_{\mu\nu}^3$. Clearly, 
for a vanishing external
field (${\overline F}_{\mu \nu }^a=0$), we have
$\varepsilon_k^2=\vec {p}^2$. 

In the case of finite temperature $T= 1/
\beta >0$, the thermodynamic potential $\Omega_q~=~-W_E/(\beta L^{3})$
\cite{5} is obtained after substituting $p_4\rightarrow \frac{2\pi
}\beta (l+\frac 12), l=0,\pm 1,\pm 2,...$,
\begin{eqnarray}
\Omega_q&=&-\frac 1{\beta L^{3}}\sum^{N_f}_{i=1}\sum_{\kappa}
\sum^{l=+\infty}_{l=-
\infty}
\sum_{k}\log\left[\left(\frac{2\pi(l+1/2)}{\beta}\right)^2+\Delta
^2+
(\eps_{k,i}-\kappa\mu)^2\right].
\label{omega}
\end{eqnarray}

Next, with the use of 
the proper time representation 
we obtain for the quark thermodynamic potential 
\begin{eqnarray}
\Omega_q&=&\frac 1{2\sqrt{\pi} L^{3}}
\sum^{N_f}_{i=1}\sum_{\kappa}
\ints^{\infty}_{1/\Lambda^2}\frac{ds}{s^{3/2}}
[1+2\sum _{l=1}^{\infty }\exp (-\frac{\beta ^2l^2}{4s}) 
(-1)^l]
\nn &&\times
\sum_{k}\exp[-s(\Delta^2+(\eps_{k,i}-\kappa\mu)^2)],
\label{munu}
\end{eqnarray}

In order to find the phase of the system at zero and finite
temperatures, solutions and energy
spectrum $\eps_{k,i}$ of the Dirac equation 
(\ref{dir}) in the chromomagnetic field should be used. For the
case of a constant abelian chromomagnetic field, equation (\ref{dir}) is decomposed into
independent equations for quarks of different colors.
Fermions are in the states determined by the following quantum
numbers: color, spin, sign of the energy, Landau number and two
components of the momentum $p_2,\,p_3$. The corresponding energy spectrum is well
known \cite{gf}: 
 \begin{eqnarray}
\eps^2_{n,\sigma,p_3}=gH(n+\frac12+
\frac{\sigma}2)+
p^2_3,
\label{14}
\end{eqnarray}
where $\sigma =\pm 1$ is the spin projection on the external field
direction, $p_3$ is the longitudinal component of the quark momentum
($-\infty
<p_3<\infty $),
\begin{eqnarray}
p_{\perp}^2&=&gH(n+\frac12)
\label{15}
\end{eqnarray}
is the transversal component squared of the quark momentum, and
$n=0,1,2,...$
is the Landau quantum number.
It should be mentioned that the boson condensate of the gluon field
$\phi$ may slightly change the structure of the fermion levels, and
besides it may mix quarks of different colors. The former may be
neglected as it will only slightly influence the total energy of
fermions, and the latter mixing of colored fermions is just the
mechanism that provides the ground state of the quark system with a
nonvanishing color charge. 
\section{CSC-CFM phase transition}
In electrodynamics, Cooper pairs can be produced only in the region,
where there is no magnetic field, as the charged condensate forces
the magnetic field outside the sample of a superconducting metal due
to the Meissner effect. In the non-abelian case the situation is
somewhat different. For instance, in the theory with the $SU(2)_c$
group, quarks in the fundamental representation form a doublet $q_i$
($i=1,2$), and the diquark condensate has the structure like
$<\varepsilon^{ij} q_i q_j>$, which is a scalar in the color space. Hence, no
contradiction between possible superconducting state creation and the
presence of a chromomagnetic field arises  \cite{Ebert}. In the real
QCD  case with the $SU(3)_c$ group, the condensate has the form
$<\epsilon^{\gamma\alpha\beta}q_{\alpha} q_{\beta}>$, where
$\alpha,\beta,\gamma=1,2,3$, and it is no more color  
neutral. This condensate 
expells the part of the color field, which
might interact with it. In the group space, a neutral chromomagnetic
field can be written in the form of a superposition of commuting
generators  $\lambda_3$ and $\lambda_8$, i.e., 
$F_{12}=B_1\lambda_3+B_2\lambda_8$. Therefore, for instance, the field
$F_{12}\sim \lambda_3$ can in principle coexist only with the
condensate of the form  $<\epsilon^{3\alpha\beta}q_{\alpha} q_{\beta}>$.

The $SU(3)_c$ group has a maximal abelian subgroup  $U(1)\times
U(1)$. In each of the subgroups  $U(1)$ either color ferromagnetism or
superconductivity is possible. In the present publication, we consider
the  $SU(2)_c$ gauge group. However, having in mind possible
generalization for the $SU(3)_c$ group, we shall consider the
situation, when color ferromagnetic and superconducting phases can
exist only separately, and a phase transition can take place between
them. 

First consider the case of zero temperature $T=0$. In order to
consider which of two phases, color ferromagnetic or supeconducting,
is preferable, one should compare the energies of these two phases. 
The ferromagnetic phase is described by a constant chromomagnetic field
${\overline F}_{12}^3=-{\overline F}_{21}^3=const=B$ and the quark
matter interacting with this 
field. The thermodynamic potential of this system is the sum of the
chromomagnetic field energy $B^2/(8\pi)$ and the total energy of
quarks $E_q$ that occupy the lowest Landau levels in this field:
\begin{equation}
E_q\,=\,\Omega_q|_{T=0,\Delta=0}\,=\,\Omega_{q 0},
\label{t0}
\end{equation}
where $\Omega_q|_{T=0,\Delta=0}$ is determined by the corresponding
limit of formula (\ref{munu}) (the first term in square brackets, and
$\Delta=0$ in the exponent). The 
contribution of gluon quantum fluctuations around the chromomagnetic field
can be neglected assuming the value of the coupling constant is small, $
g^2/(4\pi)\ll 1$, in the limit of 
larger quark density. 

In our model, the chromomagnetic field strength $B$ and its extention
$L_3$ (the dimension of the domain along the chromomagnetic field
direction) are related by the inequality (\ref{maxfield}) 
 that determines the stability condition and defines
the maximum possible chromomagnetic field strength  $B_0$
(\ref{Blength}). We will make a numerical analysis of the
thermodynamic potential
of quarks and, hence, it is 
natural to introduce dimensionless variables $\alpha,x$, and
$h(x,\alpha)$ in place of the 
chromomagnetic field  $B$, fermion density $\rho=N/L^3$ and thermodynamic
potential $\Omega_{q 0}$
\begin{equation}
 \alpha=B/B_0\,\,(\alpha\in(0,1)),\,\,x=\rho/(gB_0)^{3/2},\,\,h(x,\alpha)=\Omega_{q 0}/(2g^2B_0^2).  
\label{dimensionless}
\end{equation}   

In the absence of a chromomagnetic field the total thermodynamic
potential of fermions is
equal to 
\begin{equation}
\Omega_{q 0}|_{B=0}= 2 (gB_0)^2h(x,0).
\end{equation}
When a chromomagnetic field is generated in the system, $B=\alpha
B_0\ne 0$, the dimensionless energy of quarks becomes equal to
$h(x,\alpha)$, and the energy of the field itself should also be
added. Then
the total thermodynamic potential becomes:
\begin{equation}
\Omega_{{\rm{tot}}}\,=\,\Omega_{q0}\,+\,\frac {B^2}{8\pi}\,=\,
B_0^2\left[2g^2h(\frac{\rho}{(gB_0)^{3/2}},\alpha)\,+\,\frac{\alpha^2}{8\pi}\right].  
\end{equation}
Generation of the chromomagnetic field changes the thermodynamic potential of the
system by the amount
\begin{equation}
\frac{\Delta E}{L^3B_0^2}\,=\,\frac{\Delta \Omega}{B_0^2}\,=\,\frac{\alpha^2}{8\pi}\,+\,
2g^2\left(h(\frac{\rho}{(gB_0)^{3/2}},\alpha)-h(\frac{\rho}{(gB_0)^{3/2}},0)\right).
\end{equation}
In the model we have chosen, the chromomagnetic field  $B$ is a
variable parameter and is 
determined by the quantity  $\alpha$,
while $B_0$ and $\rho$ are assumed to be prescribed quantities. In
this case,  $\alpha$ will ''adjust'' itself in the interval  $(0,1)$
with the other parameters being fixed such that the total energy of
the system takes a minimum value,
$\Omega_{{\rm{tot}}}\to \Omega_{{\rm{min}}}$. Upon optimization with  
respect to $\alpha$,
the dimensionless energy gain due to color ferromagnetism becomes a
function only of the dimensionless fermion density $x$:
\begin{equation}
\Delta e(x)_{\rm Ferr}= \frac{\Delta \Omega_{min}}{B_0^2}\,=\, \mbox{min}_{\alpha}
\left[\frac{\alpha^2}{8\pi}\,+\,2g^2\left( h(x,\alpha)-h(x,0) \right)\right].
\end{equation}
The possible generation of a chromomagnetic field and a corresponding
ferromagnetic phase is in competion with the possible creation of a color
superconducting phase in the absence of a chromomagnetic field. The
system chooses the phase with the 
lowest 
energy. The
energy gain due to creation of a diquark condensate $\Delta \ne 0$ can
be obtained from equation (\ref{3} ) and is
equal to \cite{Rajagopal}
\begin{equation}
\Delta \Omega_{\rm CSC}=-\frac{\mu^{2}\Delta^2}{\pi^2}.
\end{equation}

Now, to decide which of the phase is
preferable, one should compare the energy gains due to production of
a color ferromagnetic 
state  $\Delta \Omega_{\rm min}$ and the superconducting state  $\Delta
\Omega_ {\rm CSC}$. For convenience, we
introduce dimensionless quantities
\begin{equation}
\begin{array}{l}
\Delta e_{\rm Ferr}\,=\,\frac{\Delta\Omega_ {\rm Ferr}}{B_0^2}\,=\,\Delta e_{\rm Ferr}(\frac{\mu^3}{6\pi^2(g B_0)^{3/2}}),
      \\
\Delta e_{\rm CSC}\,=\,\frac{\Delta
\Omega_ {\rm CSC}}{B_0^2}\,=\,-(\frac{\Delta}{\pi B_0})^2 \mu^2,\end{array}
\end{equation}
and compare them with the help of numerical calculations. 

First, the dimensionless energy gain $\Delta
h(x,\alpha)=h(x,\alpha)-h(x,0)$ 
was calculated as a function of
the dimensionless fermion density  $x$  at fixed $\alpha$. It was
demonstrated
that this difference is always negative, and this indicates that the
fermion energy is lowered when the field gets nonvanishing
values. Moreover the fermion energy 
oscillates with growing density. The reason for this is that Landau
levels are filled one by one with growing number of fermions. One may
expect that at high enough temperature these oscillations should
vanish. 

Moreover, $\Delta h(x,\alpha)$ as a function of a
dimensionless chromomagnetic field  $\alpha$ at fixed fermion density
was also calculated. It was demonstrated that the stronger the field
the lower is the 
energy of fermions until all
fermions turn out to be on the lowest Landau level. When the fermion
density is high enough, this value of 
$\alpha$ is far above the interval $(0,1)$.

In Fig. 1, the total variation of the dimensionless energy of the
whole system as a function of 
the dimensionless 
chromomagnetic field
$\alpha$ is depicted. At zero fermion density $x=0$, 
the
appearance of a
nonzero field is not favourable and the dependence is parabolic. At
finite fermion densities, 
the
growth of the chromomagnetic field is
favourable at first, but when all the fermions are on the lowest Landau
level, the energy begins to grow as $\alpha^2$. For 
higher 
fermion
density, this saturation occurs when  $\alpha>1$, i.e., the system
prefers the favourable value of  $\alpha$ close to 1 (due to
oscillations, not necessarily $\alpha=1$ ).

\begin{figure}[htbp]
\begin{center}
\renewcommand{\baselinestretch}{1}
\includegraphics[width=16cm]{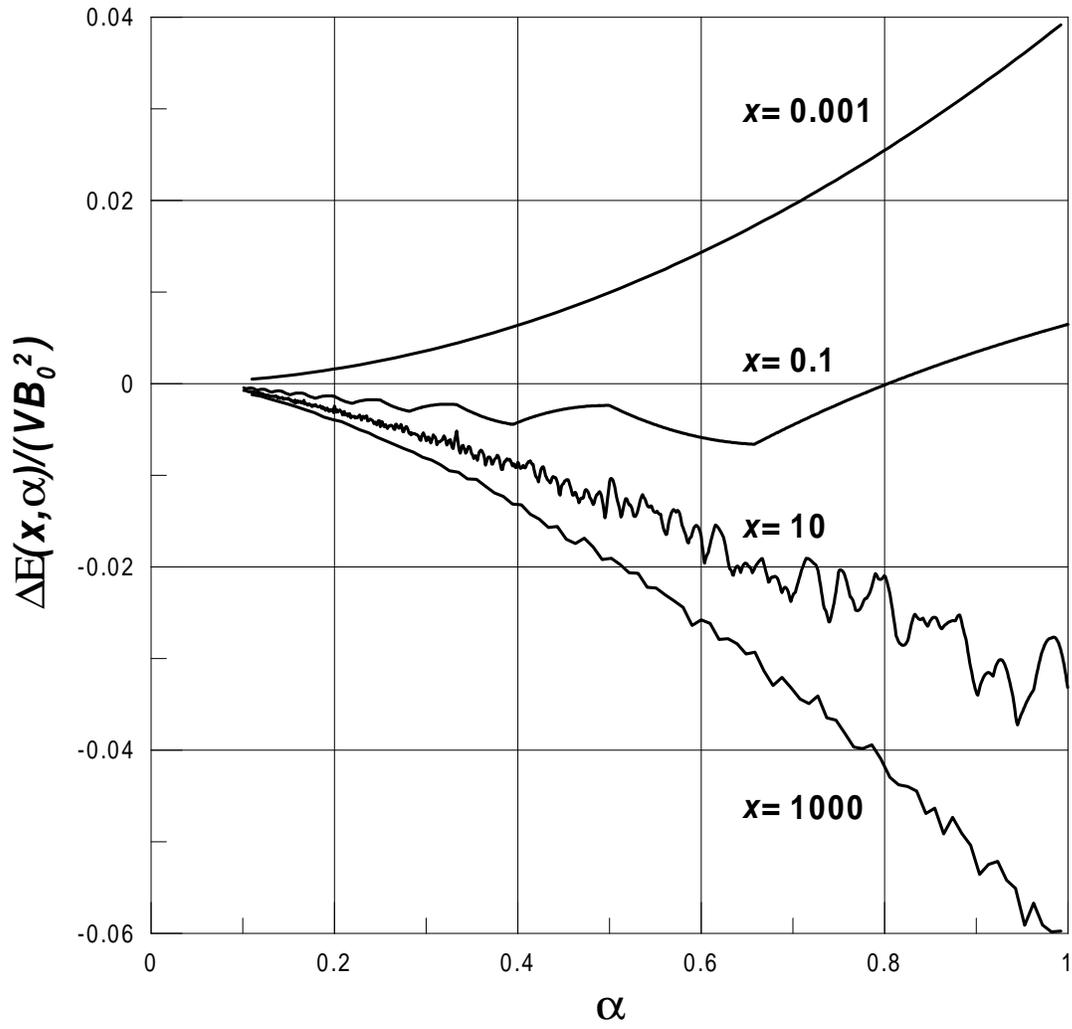}
\caption{ Total change of the dimensionless energy of the whole system as
  a function of the dimensionless chromomagnetic field  $\alpha$.} \label{fig:3}
\vspace{5mm}
\end{center}
\end{figure}

In Fig. 2, the dimensionless energy gains in color ferromagnetic  (with an
optimal value of $\alpha$ is chosen) and superconducting phases are
depicted as functions of the chemical potential  $\mu$. The choice of the
phase by the system is determined by the condition that the energy should
take 
its minimum value in this phase. 

\begin{figure}[htbp]
\begin{center}
\renewcommand{\baselinestretch}{1}
\includegraphics[width=16cm]{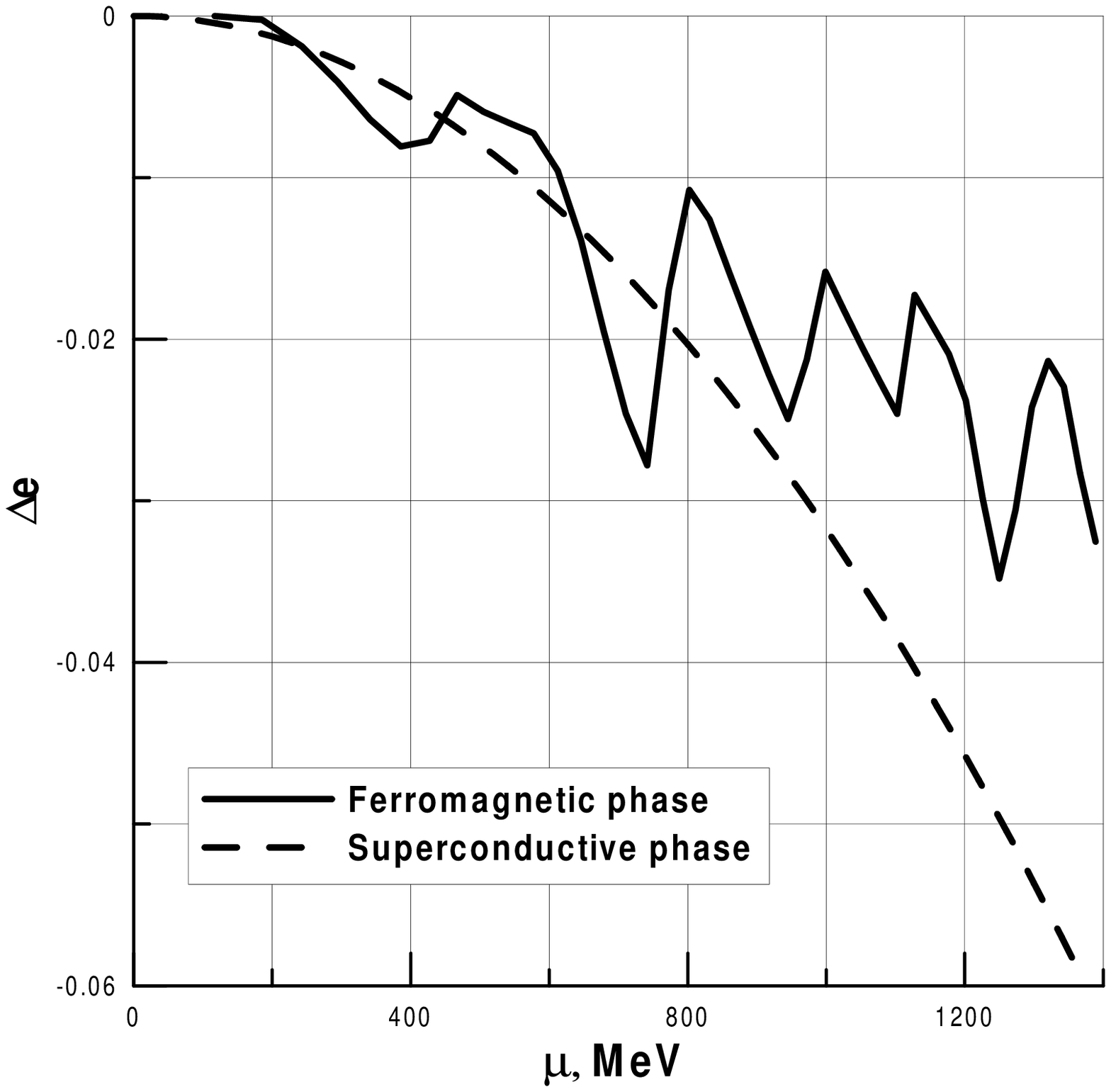}
\caption{Dimensionless energy gain in a ferromagnetic 
phase (after a choice of
  an optimal value of $\alpha$ has been done) and in a superconducting
  phase as a function of the chemical potential $\mu$.} \label{fig:4}
\vspace{5mm}
\end{center}
\end{figure}

The consideration of the phase transition in the system leads to the
conclusion that with large enough values of $\mu$, the superconducting
phase is preferable. With lowering  $\mu$, the ferromagnetic phase can
become more preferable, and a phase 
transition 
takes place. However,
one may see in the Figure that 
the energy gain corresponding to the ferromagnetic phase is an
oscillating function of $\mu$, and hence the picture of phase
transitions is somewhat more complicated. With further decrease of
$\mu$, the superconducting phase may return, then again 
the ferromagnetic
phase 
occurs, and this may repeat several times.

\section{Phase transition at finite temperature}

To describe the system at finite temperature, we should use the
formulas for the termodynamic potential of fermions
$\Omega_q(\alpha,\mu,T,B_0)$, presented in Section 2. 
Restricting our consideration only 
to the main
contribution of the ''classical '' chromomagnetic field $B^2/(8\pi)$,
and neglecting small quantum fluctuations of the gluon 
field around it, we consider the contribution of fermions, defined by
 (\ref{omega}), (\ref{munu}), where we have to put $\Delta = 0$,
and hence use the formula
\begin{equation}
\Omega_q\,=\,-T
\sum\limits_{k,i}\mbox{ln}\left[1+\mbox{exp}(\frac{\mu-\varepsilon_{k,i}}{T})\right].
\end{equation}
It should be mentioned that unlike the case of zero temperature,  at
finite temperature the chemical potential $\mu$ becomes an independent
variable and is no more equal to the Fermi energy. 

As it was done in the case of zero temperature, we employ dimensionless
variables and quantities, and moreover define a dimensionless
temperature 
\begin{equation}
\tau=\frac {T}{\sqrt{gB_0}}. 
\end{equation}
Since $\sqrt{g B_0}\sim 10^2$ MeV,  we have for the temperature  $T\sim 10^8$
eV $\sim 10^{12} K$.

Now in the case of finite temperature, we also use a numerical
calculation method, which gives only approximate results, as we have
to deal with an infinite series over the fermion energy levels. The
results are shown in Figs. 3 and 4.

In Fig. 3, the dimensionless ferromagnetic gain of the total
thermodynamic potential, minimized with respect to the dimensionless chromomagnetic
field $\alpha=B/B_0$, is depicted:
\begin{equation}
\Delta\omega(\mu, \tau)\,=\,\mbox{min}_{\alpha}\left[\frac {\alpha^2}{8\pi}\,+\,\frac
1{B_0^2}(\Omega_q(\alpha, \mu, 
T, B_0)-\Omega_q(0, \mu, T, B_0))\right].
\end{equation}
Our calculations demonstrate 
that, with growing temperature, oscillations in the
plot of $\omega$ as a function of $\mu$ become less evident and, as it
is seen in the Figure, 
in the limit of high temperature, they practically disappear. Another
interesting observation is that even at zero chemical
potential the appearance of a finite chromomagnetic field is
preferable. This is due to the fact that at nonvanishing temperature,
fermions exist with finite density even at $\mu=0$, and the appearance of
a chromomagnetic field leads to a finite energy gain. 

\begin{figure}[htbp]
\begin{center}
\renewcommand{\baselinestretch}{1}
\includegraphics[width=16cm]{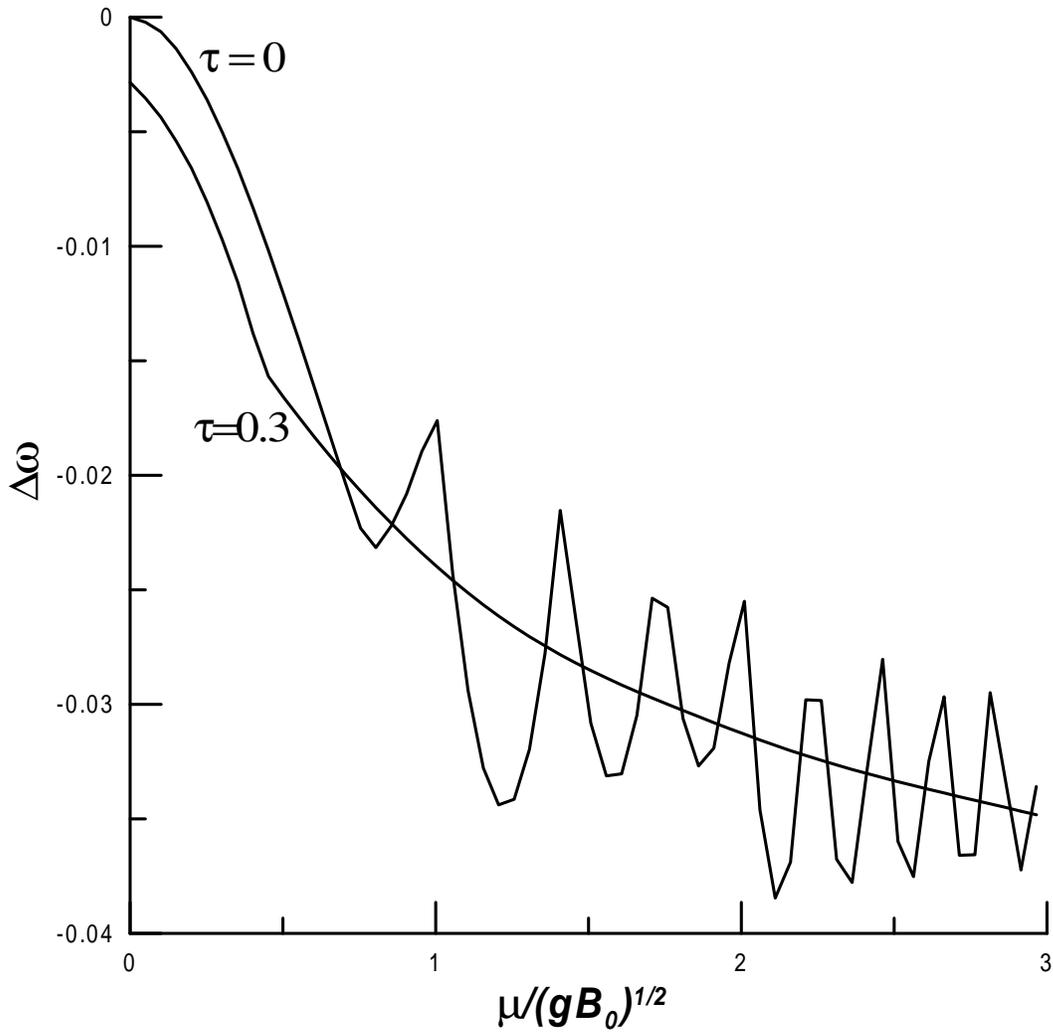}
\caption{Dimensionless thermodynamic potential gain 
$\Delta\omega$ as a
  function of  $\mu$ at zero and finite (sufficiently high)
  temperatures. The quantity $\tau$ is given by equation (30). } \label{fig:5} 
\vspace{5mm}
\end{center}
\end{figure}

Figure 4 shows that with growing temperature and at fixed  $\mu$, the
thermodynamic potential gain $\Delta\omega$ tends to a constant value.
It is seen that the higher the chemical potential, the lower temperature
$\tau(\mu)$ is needed for stabilization of $\Delta\omega$ to take
place. At low values of the chemical potential,  $\Delta\omega$ decreases
with growing temperature and it is stabilized at high enough
temperature. These facts can be explained in the following way. The
discrete 
character of the Landau levels is responsible for the
nonmonotonic behavior of $\Delta\omega(\mu)$. This situation does not
change while the temperature remains considerably lower than the
distance between levels in the vicinity of the Fermi surface. It is
clear that with growing $\mu$, the energy levels become distributed with
greater density and hence, lower temperature is needed in order to
smooth away oscillations.  

\begin{figure}[htbp]
\begin{center}
\renewcommand{\baselinestretch}{1}
\includegraphics[width=16cm]{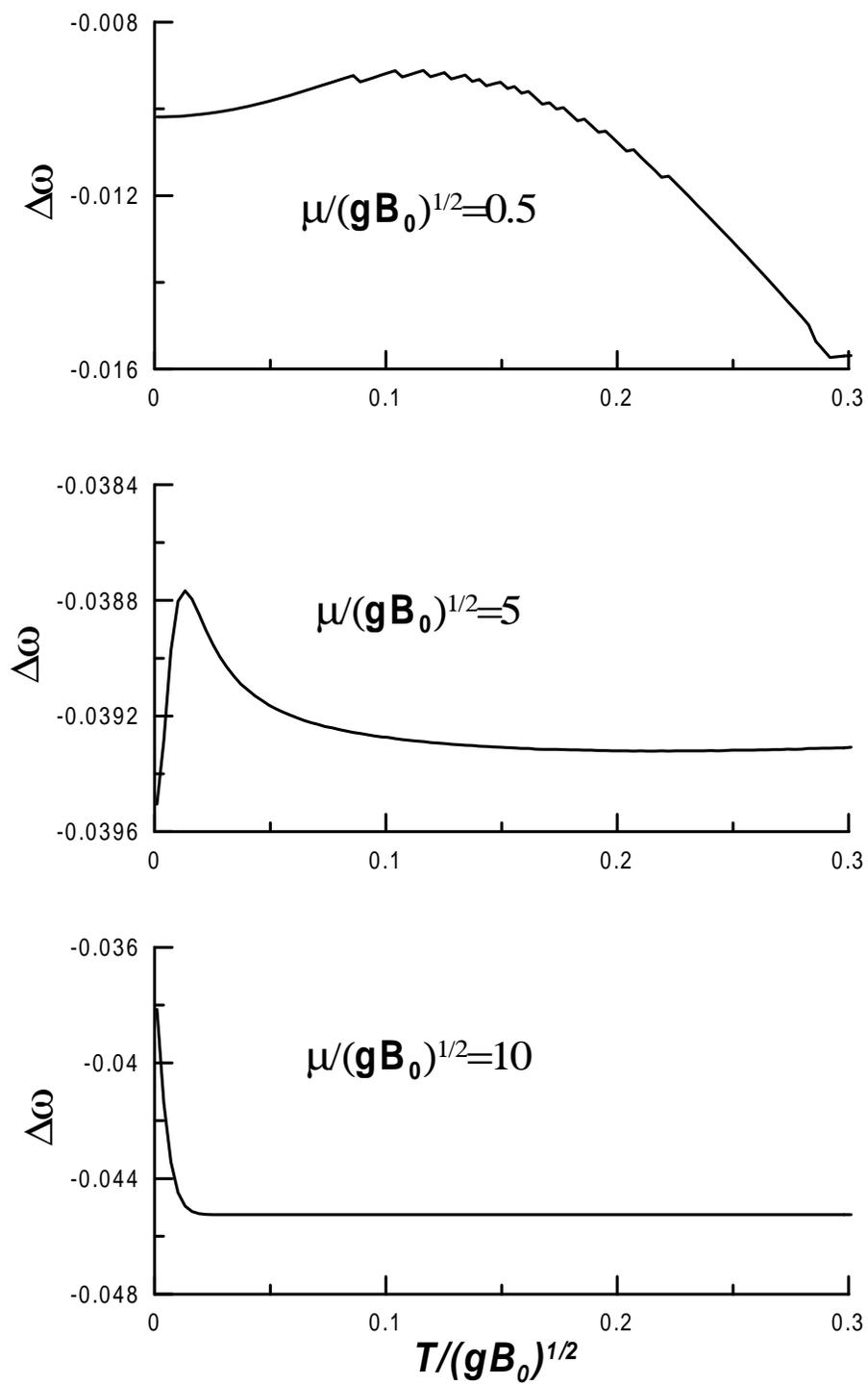}
\caption{Stabilization of the thermodynamic potential gain with
  growing  $T$ at various values of  $\mu$.} \label{fig:6} 
\vspace{5mm}
\end{center}
\end{figure}

\section{Conclusions}

In the present paper we investigated further the gauge field model
with a constant chromomagnetic field, i.e., the ferromagnetic
state. We demonstrated that the method, 
proposed in 
\cite{iwaz1}, of finding 
a
stabilized solution for this configuration,
is valid only if a physically justified condition is fulfilled, i.e., the
chromomagnetic field  exists inside certain domains with finite
dimensions. This implies that there is a maximum 
value of the chromomagnetic field inside the domain determined by the
finite spatial extentnion of the region occupied by the 
field in the direction of the field. 

The main object of our paper was to study the fermion sector of the
model, and in particular to consider  transitions between a
color superconducting phase and a possible ferromagnetic phase from
the point of view of comparing the energy gain in these phases. 
Our observations can be summarized as follows. At comparatively low
densities of the baryon matter, quarks are confined, and exist only in
colorless combinations. With growing chemical potential baryons
approach each other to such close distancies that quarks become free
particles. At sufficiently high density, the color superconducting
phase can be formed. However, for ''intermediate '' values of the
chemical potential, a ferromagnetic phase may emerge. There can be a
phase transition between these two phases depending on which phase has
larger energy gain. At zero or comparatively low temperatures a
nontrivial phase structure can be formed, and this is due to a
nonmonotonic dependence of the  energy gain of fermions in the
chromomagnetic field on their density. With growing chemical
potential, a ferromagnetic state can prove to be favorable, then it is
changed by a superconducting state, and then again ferromagnetic, and
only for high enough chemical potentials, the superconducting state
becomes always dominating. 

At sufficiently high temperatures  ($\sim \sqrt{gB_0}$), the phase
structure is simplified. The fermion energy gain is no more dependent
on temperature and becomes a monotonic function of  $\mu$. For low
$\mu$  the ferromagnetic phase becomes more favorable, and for large
enough $\mu$, the superconducting phase, there being only one
transition point between them. In the temperature scale, stabilization
occurs at  $T>T(\mu)$, where $T(\mu)$ is the minimal temperature
needed to smooth out oscillations of the thermodynamic potential
$\Omega$ that appear due to the discrete character of the fermion
spectrum in the chromomagnetic field. Theoretical reasoning and
computer calculations demonstrate that the function  $T(\mu)$ should be
decreasing. 

Further investigations will help to describe more clearly the
mechanism of formation of the chromomagnetic domains and to consider
the boundary conditions in a more sophisticated manner. 
Moreover, in order
to produce a more realistic picture of the process of the
ferromagnetic phase formation, quantum fluctuations around the ''true
vacuum '' state should be taken into account. Estimates showed that at
a chemical potential sufficiently high for the ferromagnetic state to be
formed, interactions between quarks are in no way week and should also
be considered for. Certain changes to the picture of the process may
further be given by the interaction of quarks with the charged boson condensate,
and this should also be considered. Finally, one should also
understand that in the case of the $SU(3)_c$ gauge group, there can be
states with a color ferromagnetism in 
one $U(1)$ subgroup, and
the color superconductivity in the other $U(1)$ subgroup of the
maximal Abelian subgroup $U(1)\times U(1)$.  These problems are to be
considered in further investigations. 

\section*{Acknowledgements}

Two of the authors (V.Ch.Zh. and O.V.T.) gratefully acknowledge the
hospitality of 
Prof. M.~Mueller-Preussker and his colleagues at the particle theory
group of the Humboldt University extended to them during their stay
there.
This work was partially supported by the Deutsche
Forschungsgemeinschaft under contract DFG 436 RUS 113/477/4.

\end{document}